\documentclass[lettersize,journal]{IEEEtran}
\usepackage{amsmath,amsfonts}
\usepackage{algorithmic}
\usepackage{array}
\usepackage[caption=false,font=normalsize,labelfont=sf,textfont=sf]{subfig}
\usepackage{textcomp}
\usepackage{stfloats}
\usepackage{url}
\usepackage{verbatim}
\usepackage{graphicx}
\usepackage{multirow}
\usepackage{balance}
\usepackage{cite}
\usepackage{picinpar}
\usepackage{flushend}
\usepackage[utf8]{inputenc}
\usepackage{colortbl}
\usepackage{soul}
\usepackage{pifont}
\usepackage{color}
\usepackage{alltt}
\usepackage[hidelinks]{hyperref}
\usepackage{enumerate}
\usepackage{siunitx}
\usepackage{breakurl}
\usepackage{epstopdf}
\usepackage{pbox}
\def\BibTeX{{\rm B\kern-.05em{\sc i\kern-.025em b}\kern-.08em
    T\kern-.1667em\lower.7ex\hbox{E}\kern-.125emX}}

\begin{document}

\title{G-PIFNN: A Generalizable Physics-informed Fourier Neural Network Framework for Electrical Circuits}

\author{
	\vskip 1em
	Ibrahim Shahbaz,~\IEEEmembership{Student Member,~IEEE,} Mohammad J. Abdel-Rahman,~\IEEEmembership{Senior Member,~IEEE,} and Eman Hammad,~\IEEEmembership{Senior Member,~IEEE}
	\thanks{
		I.~Shahbaz is with iSTAR Lab, Texas A\&M, College Station, TX 77843, USA. He was with the Data Science Department, Princess Sumaya University for Technology, Amman 11941, Jordan. E-mail: i.shahbaz@tamu.edu.
		M.~J.~Abdel-Rahman is with the Data Science Department, Princess Sumaya University for Technology, Amman 11941, Jordan. He is also with the Electrical and Computer Engineering Department, Virginia Tech, Blacksburg, VA 24061, USA. 
		E.~Hammad is with iSTAR Lab, Texas A\&M, College Station, TX 77843, USA.
	}
}
\maketitle

\begin{abstract}
Physics-Informed Neural Networks (PINNs) have advanced the data-driven solution of differential equations (DEs) in dynamic physical systems, yet challenges remain in explainability, scalability, and architectural complexity. This paper presents a Generalizable Physics-Informed Fourier Neural Network (G-PIFNN) framework that enhances PINN architectures for efficient and interpretable electrical circuit analysis. The proposed G-PIFNN introduces three key advancements: (1) improved performance and interpretability via a physics activation function (PAF) and a lightweight Physics-Informed Fourier Neural Network (PIFNN) architecture; (2) automated, bond graph (BG) based formulation of physics-informed loss functions for systematic differential equation generation; and (3) integration of intra-circuit and cross-circuit class transfer learning (TL) strategies, enabling unsupervised fine-tuning for rapid adaptation to varying circuit topologies. Numerical simulations demonstrate that G-PIFNN achieves significantly better predictive performance and generalization across diverse circuit classes, while significantly reducing the number of trainable parameters compared to standard PINNs.
\end{abstract}


\definecolor{limegreen}{rgb}{0.2, 0.8, 0.2}
\definecolor{forestgreen}{rgb}{0.13, 0.55, 0.13}
\definecolor{greenhtml}{rgb}{0.0, 0.5, 0.0}

\section{Introduction}

\IEEEPARstart{N}{eural} Networks (NNs) demonstrate strong learning capabilities across diverse tasks such as natural language processing, computer vision, and time series analysis~\cite{attention,CNN_foundational_paper,LSTM}. They are also widely applied in modeling physical systems including optical, mechanical, and electrical domains~\cite{DeepPhysicalNN}. Consequently, designing specialized NN architectures for surrogate modeling has become a key research priority, especially with the growing interest in realizing Digital Twins of such systems ~\cite{original_DT,AI&DT}. A major breakthrough in this direction is the development of Physics-Informed Neural Networks (PINNs)~\cite{PINNs}, a mesh-free method exploiting NNs’ expressiveness to deliver continuous solutions of differential equations DEs throughout the spatiotemporal domain~\cite{PIDL_replace_traditional_DESolvers}.

PINNs are a specialized class of NN architectures for physical modeling that integrate governing DEs as a physics-based loss term. Electrical systems have been a natural application, with studies employing PINNs from converter circuits to distribution and transmission networks. Reviews in~\cite{IEEE_ref_41,towards_PIML_power_converters} highlight their role in power systems and power electronics and provide guidelines for advancing physics-enhanced NNs beyond simple loss integration. Such improvements include designing nodes or layers to capture underlying physical processes and restructuring architectures with physical constraints, for example through customized physics activation function PAFs that enforce consistency and improve explainability.

State-of-the-art research have introduced physics-informed graphical learning architectures to build surrogate models that can generalize across minor grid topology variations~\cite{SD_ref_9,IEEE_ref_14,IEEE_ref_49,IEEE_ref_58,IEEE_ref_59}. By combining Graph NNs (GNNs) with physics-informed methods, these models capture useful feature representations of complex power systems. However, they demand massive graph datasets, rely on large number of trainable parameters, and involve lengthy offline training~\cite{GNNs_Generalization,IEEE_ref_59}. GNNs also lack interpretability, a critical drawback for safety-critical power systems applications~\cite{GNNs_Generalization}. These limitations call for scalable, efficient, and interpretable solutions that balance generalization across varying grid topologies with practical deployment~\cite{GNNs_Generalization}.

To address these limitations, recent work embeds circuit laws directly into model structure. In~\cite{IEEE_tran_OIE_5}, the introduced Physics-in-Architecture Recurrent Neural Network (PA-RNN) injects numerical forms of physical laws into a recurrent neural core, improving explainability and enabling zero-shot generalization across operating conditions. While effective for dual-active-bridge (DAB) converters, its dual-core design raises complexity and was validated on only one topology. Extending this idea, the Topology-Transferrable Physics-in-Architecture Mixture Density Network (T2PA-MDN) in~\cite{IEEE_tran_OIE_1} enables direct transfer across multiple DAB topologies—resonant, multi-level, and multi-port—without retraining, preserving physical consistency and achieving topology-aware generalization with minimal data. However, the RNN plus mixture-density components add training overhead, and the framework remains tied to the DAB family, limiting broader circuit-class generalization.

In this article, we propose a Generalizable Physics-Informed Fourier Neural Network ``G-PIFNN'' framework, which is composed of a physically enhanced PINN architecture customized for the foundational building block of electrical systems (i.e. electrical circuits). The proposed G-PIFNN framework approximates the temporal state evolution of electrical circuits in an effective and efficient manner. The improved performance is achieved by the training a lightweight Physics-Informed Fourier Neural Network (PIFNN) architecture to efficiently approximate accurate solutions of electrical quantities for diverse set of circuit topologies. The main contributions in this paper are summarized as follows:
\begin{enumerate}
    \item Proposing a tailored NN architecture for tasks related to electrical circuits simulation by augmenting PINNs with a sinusoidal physics activation function (PAF) and restructuring it to a lightweight PIFNN. This architectural synthesis has improved  the effectiveness of simulating electrical circuits with significantly lower number trainable parameters. 
    \item Introducing interpretability to NN model predictions by providing a tractable closed-form solution of electrical quantities (i.e. current and voltage).
    \item Allowing generalization for varying circuit topologies by integrating a bond graph (BG) based method in automating the physics loss term formulation and performing unsupervised fine-tuning of target models via a cross-circuit class transfer learning (TL) strategy.
\end{enumerate}

The rest of this paper is organized as follows. Section~\ref{sec: related work & Background} summarizes related work and provides preliminary background information. In Section \ref{sec: Methodology}, the methodology for constructing the physically enhanced architecture and the G-PIFNN framework is presented in detail. In Section \ref{sec: exp & results}, the designed experiments are explained and the results are reported and further discussed. Finally, Section \ref{sec: conclusion} concludes this article, providing future research directions.

\vspace{-0.1in}
\section{Related Work and Background }\label{sec: related work & Background}

\subsection{Related Work}

Physics-informed learning for power-system state estimation covers standard PINNs~\cite{IEEE_tran_TII_1}, physics-guided (PG) hybrid~\cite{IEEE_ref_18, IEEE_ref_94}, and physics-informed GNNs (PI-GNNs)~\cite{SD_ref_9, SD_ref_22} architectures. PINNs in~\cite{IEEE_tran_TII_1} demonstrated robustness to three-phase faults and false-data-injection versus purely data-driven NNs. PG models~\cite{IEEE_ref_18, IEEE_ref_94} use stacked auto-encoders to map measurements to states and reconstruct measurements, but at substantial parameter complexity. PI-GNNs~\cite{SD_ref_9, SD_ref_22} generalize to minor topology changes, yet require extensive data and computation. Despite advances, these methods remain black-box with limited explainability, constraining validation and trust in safety-critical power-system applications.

Physics-informed learning has also been applied to power flow analysis (PFA) and optimal power flow (OPF), including PG parameter initialization~\cite{IEEE_ref_81} and KKT-based incorporation~\cite{SD_ref_6, SD_ref_13}. A novel physics-informed Jacobian-tensor gradient estimator method for alternating current (AC) AC-OPF solutions is introduced in ~\cite{IEEE_tran_TII_2}, which preserves ground-truth convergence with far lower runtime and storage, while graph models—graph convolutional NNs (GCNNs)~\cite{IEEE_ref_14, IEEE_ref_49} and graph attention networks (GATs)~\cite{IEEE_ref_23}—encode grid topology; nevertheless, these methods still suffer from data dependence, computational cost, limited interpretability, and limited generalization across varying grid topologies.

Physics-informed approaches for power-system stability include physics-informed features extraction with ensemble Gaussian processes for voltage-stability margins prediction~\cite{IEEE_ref_87}, PG-NNs that improve load-margin prediction over standard NNs~\cite{IEEE_ref_32}, and a physics-augmented auxiliary-learning framework for transient stability employing a multi-gate mixture-of-experts with auxiliary electrical-velocity prediction and dual-phase PG training~\cite{IEEE_tran_TII_3}. Despite accuracy gains, these methods still rely on labeled data for feature extraction, require long training for complex PG models, and provide limited model interpretability.

Standard PINN architectures have been deployed in modeling power electronics devices~\cite{IEEE_ref_3,IEEE_ref_5,IEEE_ref_35,IEEE_tran_OIE_2}. Applications include estimating unknown circuit parameters in buck converters~\cite{IEEE_ref_3}, identifying the online impedance of voltage source converters (VSCs)~\cite{IEEE_ref_35,IEEE_tran_OIE_2}, and modeling grid-tied converters under disturbances~\cite{IEEE_ref_5}. Current limitations of the deployed standard PINNs architecture include the high model complexity, inability to generalize to varying converter topologies and lack of model predictions interpretability.

\vspace{-0.1in}

\subsection{Background}
\begin{figure*}[!t]
\centering
\subfloat[]{\includegraphics[width=1.8in]{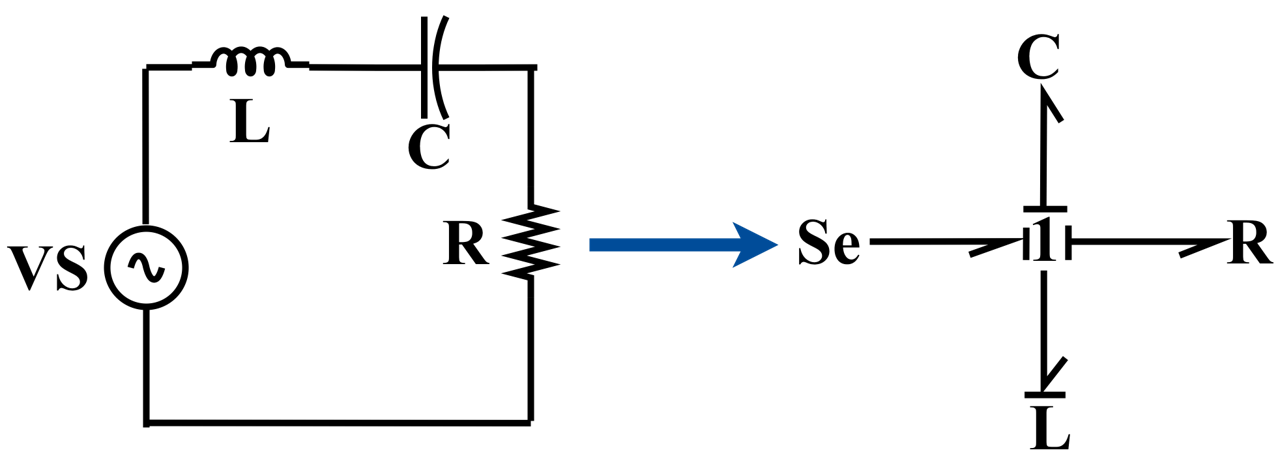}%
\label{Fig1_a}}
\hfil
\subfloat[]{\includegraphics[width=2.3in]{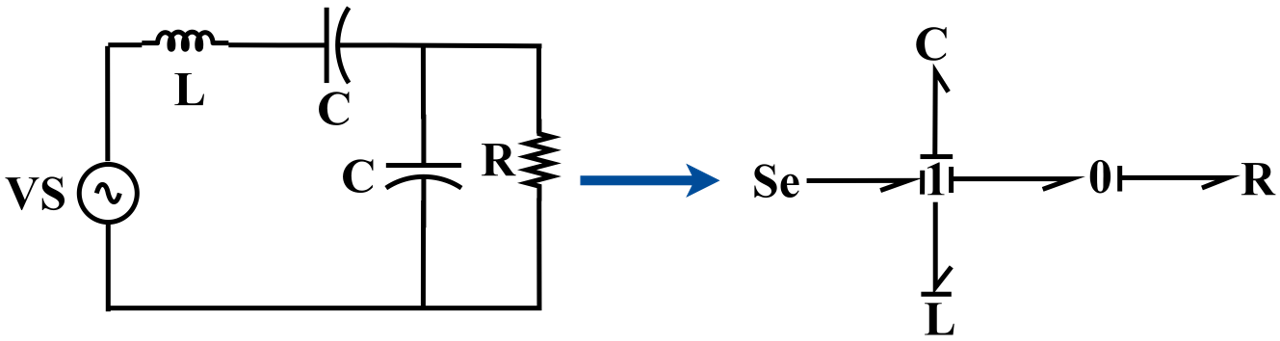}%
\label{Fig1_b}}
\hfil
\subfloat[]{\includegraphics[width=3in]{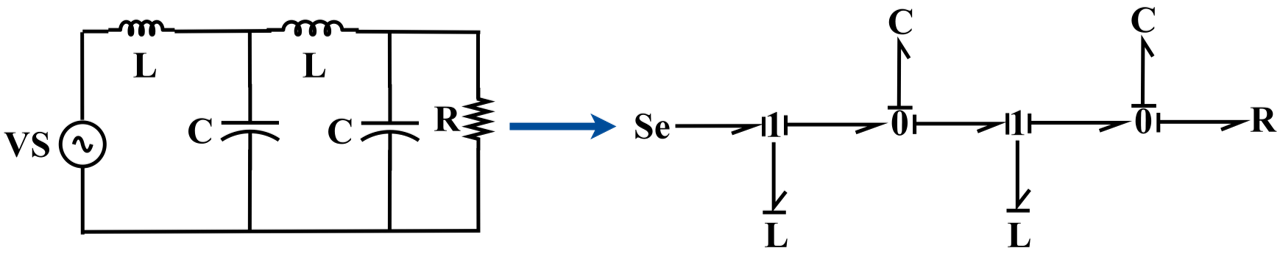}%
\label{Fig1_c}}
\hfil
\caption{Bond graph modeling for (a) a class 1 circuit, (b) a class 2 circuit, and (c) a class 3 circuit.}
\label{Fig1}
\end{figure*}

BGs provide a unified graphical approach for representing energy transfer in dynamical systems across mechanical, electrical, hydraulic, and thermal domains. Based on energy conservation and system analogies, they offer a systematic framework for modeling energy flow~\cite{BGs_Paynter}. In electrical systems, BGs use components such as sources ($Se$ for voltage, $Sf$ for current), detectors ($De$, $Df$), resistors ($R$), capacitors ($C$), and inductors ($L$). Components connect through parallel $0$ junctions and series $1$ junctions. Fig.~\ref{Fig1} shows the conversion of three classes of electrical circuits into BG representation. BG modeling has been employed to supply GNNs with modular graph-based circuit representations for accurate state estimation in power electronics~\cite{khamis2024circuit}. Moreover, the automated method in~\cite{electrical_system_simulation_via_BGs} to transform BGs into DEs is adopted in this work as a reliable DEs derivation approach to support the generalization need for varying circuit topologies.

\section{Methodology}\label{sec: Methodology}

\begin{figure*}[!t]
\centering
\includegraphics[width=\textwidth]{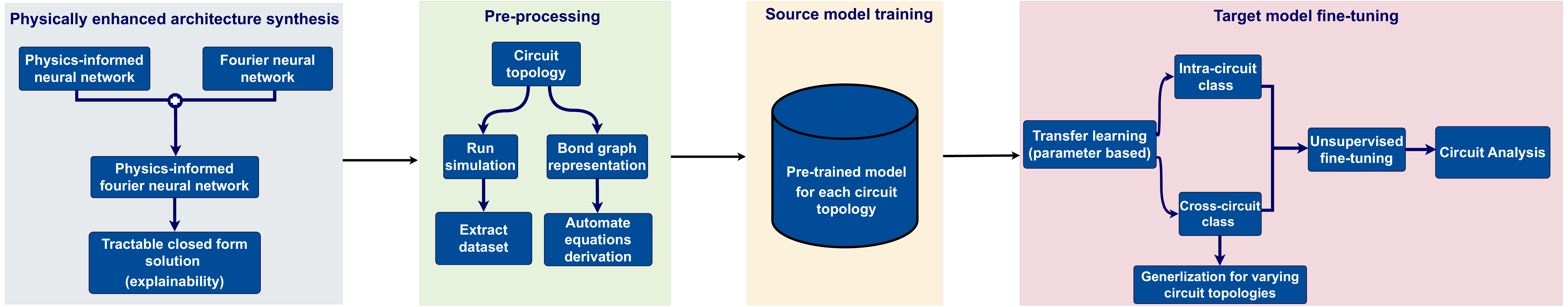}
\caption{The G-PIFNN framework.}
\label{Fig2}
\end{figure*}

In this section, we assemble our proposed physically enhanced PINN architecture, PIFNN, and present the G-PIFNN framework. The overall G-PIFNN framework is illustrated in Fig.~\ref{Fig2} and is composed of four main consecutive phases: physically enhanced architecture synthesis, pre-processing, source model training, and target model fine-tuning. Each phase is described in the following subsections.

\subsection{Physically Enhanced Architecture Synthesis}

The PIFNN design is motivated by the sinusoidal nature of AC signals, and is achieved by incorporating a smooth cosine PAF into the PINN architecture~\cite{PIFNN}. This activation enforces physical consistency with sinusoidal outputs but does not improve training efficiency, maintaining baseline PINN performance. To address this, the deep multilayer NN in a standard PINN is replaced with a shallow FNN, forming the PIFNN architecture. The improved efficiency is demonstrated through reduced model complexity, which is realized by the decreased number of trainable network parameters, due to synthesizing the shallow PIFNN architecture.

The assembled PIFNN, shown in Fig.~\ref{Fig3}, is designed to approximate temporal solutions of electrical systems. The architecture enhances interpretability  by yielding  a tractable closed-form expression for the monitored electrical quantity and its derivatives. For a PIFNN with $N$ neurons, the solution for $I(t)$ takes the Fourier-like form $I(t)=\alpha_{0}+\sum_{k=1}^{N}\lambda_{k}\cos(W_{k}t+b_{k})$, where the $n$th derivative of the same quantity can be derived as follows:
\begin{equation}\label{eqn:Closed form Solution for d^n I(t)/dt^n}
\frac{\partial^n I(t)}{\partial t^n} = 
\begin{cases}
(-1)^{n/2} \sum_{k=1}^{N} \lambda_{k} W_{k}^n \cos(W_{k} t + b_{k}); n \text{ is even} \\
(-1)^{\lceil n/2 \rceil} \sum_{k=1}^{N} \lambda_{k} W_{k}^n \sin(W_{k} t + b_{k}); n \text{ is odd}.
\end{cases}
\end{equation}

In terms of the network's trainable parameters, 
$\theta = \{W_{1}, \ldots, W_{N}, b_{1}, \ldots, b_{N}, \lambda_{1}, \ldots, \lambda_{N}, \alpha_{0}\}$, 
the general physics-informed loss function $L(\theta,\Phi)$ of the PIFNN is defined as the sum of three Mean Squared Error (MSE) terms:
\begin{equation}\label{eqn:PIFNN Loss Function}
L(\theta,\Phi) = L_{\text{data}}(\theta) + L_{\text{pde}}(\theta,\Phi) + L_{\text{ic}}(\theta),
\end{equation}
where $\Phi = \{R, L, C, V_{\max}, f\}$ are the circuit parameters: resistance, inductance, capacitance, maximum voltage amplitude, and source frequency. 
Here, $L_{\text{data}}(\theta)$ represents the supervised data loss, 
$L_{\text{pde}}(\theta,\Phi)$ ensures the network output satisfies the governing circuit DEs, 
and $L_{\text{ic}}(\theta)$ enforces the initial conditions of the system.

Although this study focuses on AC RLC circuits, the methodology readily extends to broader parameter spaces for approximating signals in power-system and power-electronics applications. Fourier cosine series approximate not only sinusoidal but also piecewise-smooth and discontinuous waveforms by meeting Dirichlet conditions for periodic functions with finite discontinuities and bounded variation~\cite{tolstov2012fourier}. Under \(L^2\)-convergence~\cite{katznelson2004introduction}, cosine expansions represent integrable functions, including those with complex amplitude–phase structure. Additionally, Fourier series can decompose triangular and rectangular pulses into harmonics whose superposition synthesizes diverse signals~\cite{bracewell1966fourier}.

The proposed PIFNN architecture can be configured to solve both forward and inverse problems by minimizing the loss formulated in~\eqref{eqn:PIFNN Loss Function}. In a forward solution mode, the optimizer is configured to only optimize NN parameters $\theta$. On the other hand, in an inverse solution mode, the optimizer is configured to jointly optimize the set of parameters \{$\theta$,$\Phi$\}.

\subsection{Pre-processing}

In the pre-processing phase, the dataset is extracted, and the electrical system equations are derived. To generate the dataset, a simulation tool is first used to simulate the temporal state evolution of a certain circuit topology while monitoring the physical quantities of interest (i.e., current or voltage). The dataset is then extracted in a tabular format and stored in a Comma Separated Values (CSV) file.

The selected circuit classes evaluate the framework's ability to solve and generalize between simple (low-order DEs) and complex (high-order DEs) circuits. The BG method, previously tested, was not implemented in the following experiments, and its DEs derivation overhead is excluded. It is suggested here as a reliable equation-derivation method to support the framework's generalization across varying circuit topologies.

A reliable equation-derivation method is employed to automate generation of the DEs governing the circuit topology. This stage highlights the need for such automation to formulate the physics loss term in a physics-informed loss function. Using the BG-based approach in~\cite{electrical_system_simulation_via_BGs}, we can derive the following set of non-homogeneous DEs for circuit classes 1, 2, and 3 shown in Fig.~\ref{Fig1}:
\begin{align}
\frac{\partial V_{S}(t)}{\partial t} & = 
L \; \frac{\partial^2 I_{load}(t)}{\partial t^2} + R_{load} \; \frac{\partial I_{load}(t)}{\partial t} \nonumber \\
& \hspace{1.9in} + \frac{1}{C} \; I_{load}(t) \label{eqn:Class_1_DE} \\
\frac{\partial V_{S}(t)}{\partial t} & = R_{load} L C \;\frac{\partial^3 I_{load}(t)}{\partial t^3} + L \; \frac{\partial^2 I_{load}(t)}{\partial t^2} \nonumber \\
& \hspace{0.75in} + 2 R_{load} \; \frac{\partial I_{load}(t)}{\partial t} + \frac{1}{C} \; I_{load}(t) \label{eqn:Class_2_DE} \\
V_{S}(t) & =
R_{load} L^2 C^2 \; \frac{\partial^4 I_{load}(t)}{\partial t^4} + 3 L C \; \frac{\partial^2 I_{load}(t)}{\partial t^2} \nonumber \\
& \hspace{0.25in} + (L^2 C + 2L) \; \frac{\partial I_{load}(t)}{\partial t}+ R_{load} \; I_{load}(t),\label{eqn:Class_3_DE}
\end{align}
where $I_{load}$ is the current through $R=R_{load}$, and $V_{S}(t) = V_{\max} \sin(\omega t)$ denotes a sinusoidal voltage source with amplitude $V_{\max}$. Here, $\omega = 2\pi f$ is the angular frequency (rad/seconds), and $f$ the cyclic frequency (Hz). The DEs~\eqref{eqn:Class_2_DE}-\eqref{eqn:Class_3_DE} are derived under the assumption of equal inductance and capacitance in class 2 and class 3 circuits.

\newcommand{\figVscale}{0.7} 
\begin{figure*}[!t]
\centering
\scalebox{1}[\figVscale]{%
  \includegraphics[width=\textwidth]{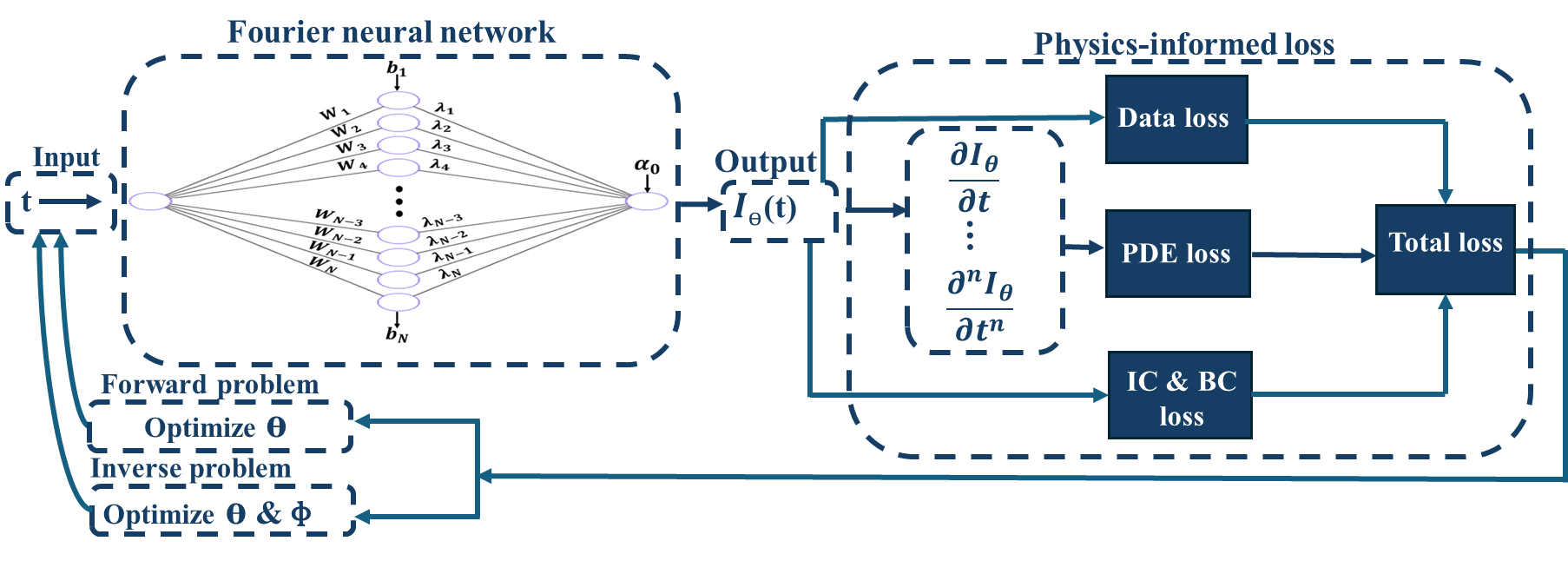}%
}
\caption{Assembling a PIFNN architecture for forward and inverse mode solutions.}
\label{Fig3}
\end{figure*}

\subsection{Source Model Training}

This phase pre-trains the source PIFNNs. After configuring the PIFNN for forward problems, the source model \(\text{PIFNN}_{s}\) is trained on simulated data and derived DEs from pre-processing. Each circuit topology receives a dedicated \(\text{PIFNN}_{s}\) for subsequent transfer learning. Training comprises four steps: parameter initialization, physics-informed loss design, forward-mode optimization, and saving the best parameters.

A systematic initialization of PIFNN$_{s}$ parameters, $\theta_{s}$, is employed to prevent vanishing/exploding gradients and accelerate training convergence. Following~\cite{PIFNN}, all biases ($b_{1, \text{init}}, \ldots, b_{N, \text{init}}, \alpha_{0, \text{init}}$) are set to zero. Input-to-hidden layer weights are drawn from a normal distribution with mean 0 and variance 5, while hidden-to-output layer weights are drawn from another normal distribution with mean 0 and variance $0.9703/N$, ensuring equal variance across input-output layers. This calibration stabilizes training and accelerates convergence, as demonstrated theoretically and experimentally in~\cite{PIFNN}.

After initializing PIFNN$_{s}$ parameters, the physics-informed loss function is set up as in~\eqref{eqn:PIFNN Loss Function}, comprising data, physics, and initial condition loss terms. The optimal parameters, $\theta_{s, \text{opt}}$, are obtained by training the forward-mode PIFNN$_{s}$ to iteratively minimize this loss. Once trained, $\theta_{s, \text{opt}}$ are saved for subsequent target model fine-tuning phase.

\subsection{Target Model Fine-tuning}

In the final phase, we perform electric circuit analysis, where the goal is to approximate an effective solution for the electrical quantity being analyzed in the circuit given new circuit parameter values, $\Phi_{\text{analysis}}$.

In the first step, TL is performed between  PIFNN$_{s}$ and a target PIFNN. The knowledge transfer between both models takes the form of parameter-based TL, where the previously saved $\theta_{s, \text{opt}}$ parameters are used to initialize the target PIFNN model parameters $\theta_{t, \text{analysis}}$. Hence, both source and target PIFNN models must share the same network architecture, i.e., the same number of layers and neurons in each layer. 

We propose two parameter-based TL strategies to achieve generalization for target circuit analysis tasks: intra-circuit class TL and cross-circuit class TL. In intra-circuit class TL, both source and target PIFNN models share the same circuit topology. On the other hand, in cross-circuit class TL, the target PIFNN model learning task is associated with a different circuit topology from PIFNN$_{s}$learning task circuit topology. The cross-circuit class TL strategy is proposed to prove the effective and efficient generalization of the G-PIFNN framework to varying circuit topologies.

After performing TL, the physics-informed loss function is updated to account for circuit parameters and/or topology changes. In the following two steps, the user then selects new circuit parameter values ($\Phi_{\text{analysis}}$) for the circuit chosen topology under investigation, and updates the physics-informed loss function accordingly. The physics-informed loss function for the target model ($L_{t, \text{analysis}}$) is constructed without a data term and is only composed of the physics and the initial conditions loss terms. Its mathematical formulation is as follows:
\begin{equation}\label{eqn: Analysis Target PIFNN Loss Function}
L_{t, \text{analysis}}(\theta, \Phi_{\text{analysis}}) = L_{\text{pde}}(\theta, \Phi_{\text{analysis}}) + L_{ic}(\theta).
\end{equation}
This formulation enables the user to simulate new solutions for the selected circuit topology under investigation in an unsupervised manner.

\begin{figure*}[!t]
\centering
\begingroup
\renewcommand{\figVscale}{0.6}
\scalebox{1}[\figVscale]{%
  \includegraphics[width=\textwidth]{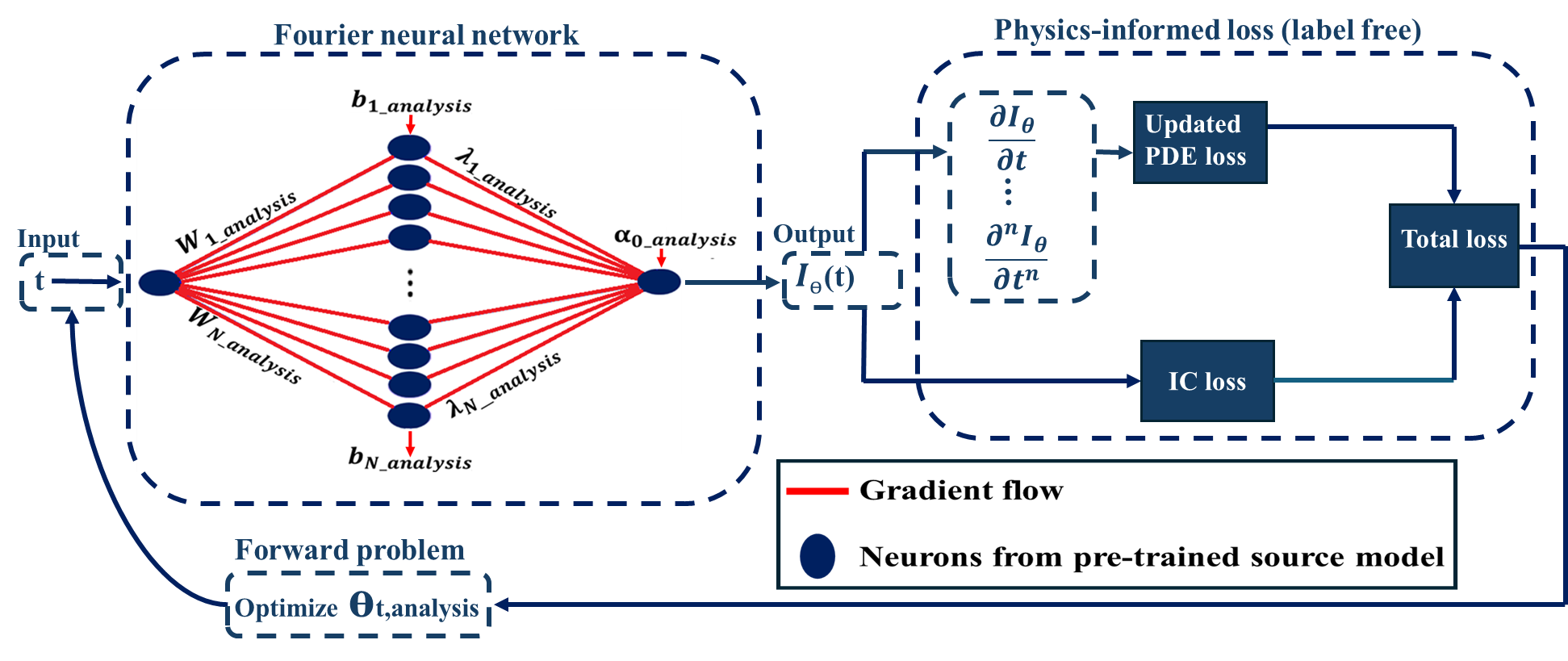}%
}
\caption{Performing circuit analysis through target model unsupervised fine-tuning.}
\label{Fig4}
\endgroup
\end{figure*}

The final step of this phase is illustrated in Fig.~\ref{Fig4}. A forward mode configured target PIFNN model is trained to iteratively minimize the physics-informed loss function formulated in~\eqref{eqn: Analysis Target PIFNN Loss Function}. The training process is an unsupervised fine-tuning process, where the training is guided by physical knowledge only, mitigating the need for labeled data.

\section{Experimentation and Results}\label{sec: exp & results}

In this section, we discuss the general experimentation setup, the designed experiments, and the results that verify the effectiveness, efficiency, and generalizability of the proposed G-PIFNN framework. Our PyTorch implementation of G-PIFNN is
available at https://github.com/i-shahbaz-tamu/G-PIFNN.git.

Experiments are performed on a personal laptop with an Intel Core i5-9300H CPU at $2.40$~GHz, $32$~GB RAM, and an NVIDIA GeForce GTX~1050 GPU. The software code is implemented in PyTorch~v2.3.1+cu121, while Simulink (MATLAB R2024a) software was used to extract ground-truth data and compare results. For benchmarking, the standard PINN architecture and training setup from~\cite{IEEE_ref_3} is reproduced for PINN models. PIFNN and PINN models are trained using the ADAM optimizer~\cite{ADAM} with varying learning rates.

All experiments involve circuit analysis on the three electrical circuit classes shown in Fig.~\ref{Fig1}, selected to evaluate the framework's ability to solve forward problems with varying DE orders. The target quantity in all circuits is the load current through $R_{\text{load}}$, with DEs~\eqref{eqn:Class_1_DE}-\eqref{eqn:Class_3_DE} incorporated in the physics-informed loss.  

Simulation data is extracted in CSV format and converted to tensor-based training and testing datasets, randomly split $50\%$ for training and $50\%$ for testing. The simulation runs are performed for the temporal intervals between 0 and 1 seconds, sufficient to capture the dynamics and periodicity of the signals. Model predictions are visualized over the temporal interval $t = [0.5, 1]$ seconds of the testing split.

\subsection{Source Model Training Experiment}

In the first experiment, simulations were conducted to monitor the load current in the three circuit classes, and source PIFNN models were trained using the extracted data and embedded physical knowledge. The experiment aims to compare the performance of the source PIFNN against standard PINNs and simulation results in terms of accuracy and efficiency. Initial parameters for each circuit class are listed in Table~\ref{Table1}. Two simulations per circuit class were run for 0.5 seconds duration each: the first over $t=[0,0.5]$ seconds and the second over $t=[0.5,1]$ seconds.

\begin{table}[]
\begin{center}
\caption{Initial circuit parameter settings.}
\label{Table1}
\begin{tabular}{|>{\centering\arraybackslash}p{0.12\linewidth}|>{\centering\arraybackslash}p{0.11\linewidth}|>{\centering\arraybackslash}p{0.11\linewidth}|>{\centering\arraybackslash}p{0.11\linewidth}|>{\centering\arraybackslash}p{0.12\linewidth}|>{\centering\arraybackslash}p{0.12\linewidth}|}
\hline
\textbf{Circuit Class} & \textbf{$R$} (Ohms) & \textbf{$L$} (Henry) & \textbf{$C$} (Farads) & \textbf{$V_{\max}$} (Volts) & \textbf{$f$} (Hertz) \\ \hline
\textbf{Class 1} & $5$  & $0.005$ & $0.009$   & $10$  & $30$ \\ \hline
\textbf{Class 2} & $50$ & $0.001$ & $0.00009$ & $150$ & $30$ \\ \hline
\textbf{Class 3} & $50$ & $0.005$ & $0.00006$ & $20$  & $30$ \\ \hline
\end{tabular}
\end{center}
\end{table}

The PIFNN models are subjected to hyper-parameter tuning to determine the optimal number of neurons in the single-hidden-layer architecture, $N_{\text{neurons}}$, with values explored in $[10, 20, 30, 40, 50]$. Training runs for 600 epochs with a two-stage schedule: learning rate \(10\) for the first 300 epochs to accelerate progress, then \(10^{-3}\) for the remaining 300 epochs to refine convergence. For comparison, the PINN in~\cite{IEEE_ref_3} uses a tanh multi-layer network with 50 neurons per layer; we tune \(N_{\text{hidden}}\in\{1,2,3,4,5\}\) and train for 100{,}000 epochs at a constant learning rate \(10^{-3}\).

Separate PIFNN and PINN models are trained and tested for each circuit class under different hyperparameter settings. The effectiveness of each model is measured using the MSE metric. The reported MSE is measured between the model's prediction and actual simulation ground-truth data results. The efficiency of each model is measured in terms of training and testing duration. The training and testing results of this experiment are summarized in Table~\ref{Table2}, and the simulation execution duration for each circuit class is reported in Table~\ref{Table3}.
\begin{table*}[]
\begin{center}
\caption{Results of simulation comparison between PIFNN and PINN:  Source model training and testing.}
\label{Table2}
\resizebox{\textwidth}{!}{%
\begin{tabular}{|c|ccccc|ccccc|}
\hline
\multirow{2}{*}{\textbf{Circuit Class}} &
  \multicolumn{5}{c|}{\textbf{PIFNN}} &
  \multicolumn{5}{c|}{\textbf{PINN}} \\ \cline{2-11} 
 &
  \multicolumn{1}{c|}{\textbf{$N_{\text{neurons}}$}} &
  \multicolumn{1}{c|}{\textbf{\begin{tabular}[c]{@{}c@{}}Training\\ MSE\end{tabular}}} &
  \multicolumn{1}{c|}{\textbf{\begin{tabular}[c]{@{}c@{}}Training \\ Duration\\ (seconds)\end{tabular}}} &
  \multicolumn{1}{c|}{\textbf{\begin{tabular}[c]{@{}c@{}}Testing \\ MSE\end{tabular}}} &
  \textbf{\begin{tabular}[c]{@{}c@{}}Testing \\ Duration\\ (seconds)\end{tabular}} &
  \multicolumn{1}{c|}{\textbf{$N_{\text{hidden}}$}} &
  \multicolumn{1}{c|}{\textbf{\begin{tabular}[c]{@{}c@{}}Training \\ MSE\end{tabular}}} &
  \multicolumn{1}{c|}{\textbf{\begin{tabular}[c]{@{}c@{}}Training \\ Duration\\ (minutes)\end{tabular}}} &
  \multicolumn{1}{c|}{\textbf{\begin{tabular}[c]{@{}c@{}}Testing \\ MSE\end{tabular}}} &
  \textbf{\begin{tabular}[c]{@{}c@{}}Testing \\ Duration\\ (seconds)\end{tabular}} \\ \hline
\multirow{5}{*}{\textbf{Class 1}} &
  \multicolumn{1}{c|}{$10$} &
  \multicolumn{1}{c|}{$2.76000E-03$} &
  \multicolumn{1}{c|}{$1.42$} &
  \multicolumn{1}{c|}{$1.51150E-05$} &
  $0.00053$ &
  \multicolumn{1}{c|}{$1$} &
  \multicolumn{1}{c|}{$2.79004E+00$} &
  \multicolumn{1}{c|}{$9.73017$} &
  \multicolumn{1}{c|}{$2.79004E+00$} &
  $0.00049$ \\ \cline{2-11} 
 &
  \multicolumn{1}{c|}{$20$} &
  \multicolumn{1}{c|}{$2.26000E-03$} &
  \multicolumn{1}{c|}{$1.38$} &
  \multicolumn{1}{c|}{$1.33560E-01$} &
  $0.00026$ &
  \multicolumn{1}{c|}{$2$} &
  \multicolumn{1}{c|}{$4.62414E+02$} &
  \multicolumn{1}{c|}{$13.58613$} &
  \multicolumn{1}{c|}{$4.62414E+02$} &
  $0.00056$ \\ \cline{2-11} 
 &
  \multicolumn{1}{c|}{$30$} &
  \multicolumn{1}{c|}{$4.15000E-03$} &
  \multicolumn{1}{c|}{$1.41$} &
  \multicolumn{1}{c|}{$3.80588E+00$} &
  $0.00041$ &
  \multicolumn{1}{c|}{$3$} &
  \multicolumn{1}{c|}{$1.66445E+01$} &
  \multicolumn{1}{c|}{$15.25606$} &
  \multicolumn{1}{c|}{$1.66445E+01$} &
  $0.00078$ \\ \cline{2-11} 
 &
  \multicolumn{1}{c|}{$40$} &
  \multicolumn{1}{c|}{$3.20510E-01$} &
  \multicolumn{1}{c|}{$1.40$} &
  \multicolumn{1}{c|}{$4.25777E+01$} &
  $0.00036$ &
  \multicolumn{1}{c|}{$4$} &
  \multicolumn{1}{c|}{$1.23774E+01$} &
  \multicolumn{1}{c|}{$19.8105$} &
  \multicolumn{1}{c|}{$1.23774E+01$} &
  $0.00094$ \\ \cline{2-11} 
 &
  \multicolumn{1}{c|}{$50$} &
  \multicolumn{1}{c|}{$1.47500E-02$} &
  \multicolumn{1}{c|}{$1.51$} &
  \multicolumn{1}{c|}{$3.64345E+00$} &
  $0.00037$ &
  \multicolumn{1}{c|}{$5$} &
  \multicolumn{1}{c|}{$1.04476E+00$} &
  \multicolumn{1}{c|}{$23.59783$} &
  \multicolumn{1}{c|}{$1.04476E+01$} &
 $0.00106$ \\ \hline
\multirow{5}{*}{\textbf{Class 2}} &
  \multicolumn{1}{c|}{$10$} &
  \multicolumn{1}{c|}{$7.16000E-03$} &
  \multicolumn{1}{c|}{$1.64$} &
  \multicolumn{1}{c|}{$7.92000E-03$} &
  $0.00030$ &
  \multicolumn{1}{c|}{$1$} &
  \multicolumn{1}{c|}{$8.78627E-01$} &
  \multicolumn{1}{c|}{$17.14341$} &
  \multicolumn{1}{c|}{$8.78627E-01$} &
  $0.00047$ \\ \cline{2-11} 
 &
  \multicolumn{1}{c|}{$20$} &
  \multicolumn{1}{c|}{$7.49000E-03$} &
  \multicolumn{1}{c|}{$1.78$} &
  \multicolumn{1}{c|}{$1.05000E-03$} &
  $0.00033$ &
  \multicolumn{1}{c|}{$2$} &
  \multicolumn{1}{c|}{$4.89093E-01$} &
  \multicolumn{1}{c|}{$26.63067$} &
  \multicolumn{1}{c|}{$4.89092E-01$} &
  $0.00067$ \\ \cline{2-11} 
 &
  \multicolumn{1}{c|}{$30$} &
  \multicolumn{1}{c|}{$1.03200E-02$} &
  \multicolumn{1}{c|}{$1.84$} &
  \multicolumn{1}{c|}{$1.97907E+00$} &
  $0.00026$ &
  \multicolumn{1}{c|}{$3$} &
  \multicolumn{1}{c|}{$2.81396E-01$} &
  \multicolumn{1}{c|}{$40.16972$} &
  \multicolumn{1}{c|}{$2.81396E-01$} &
  $0.00072$ \\ \cline{2-11} 
 &
  \multicolumn{1}{c|}{$40$} &
  \multicolumn{1}{c|}{$2.89400E-02$} &
  \multicolumn{1}{c|}{$1.95$} &
  \multicolumn{1}{c|}{$3.35589E+00$} &
  $0.00033$ &
  \multicolumn{1}{c|}{$4$} &
  \multicolumn{1}{c|}{$3.45231E-01$} &
  \multicolumn{1}{c|}{$53.24517$} &
  \multicolumn{1}{c|}{$3.45231E-01$} &
  $0.00099$ \\ \cline{2-11} 
 &
  \multicolumn{1}{c|}{$50$} &
  \multicolumn{1}{c|}{$2.45000E-02$} &
  \multicolumn{1}{c|}{$1.98$} &
  \multicolumn{1}{c|}{$1.81342E+01$} &
  $0.00033$ &
  \multicolumn{1}{c|}{$5$} &
  \multicolumn{1}{c|}{$1.50754E-01$} &
  \multicolumn{1}{c|}{$76.71552$} &
  \multicolumn{1}{c|}{$1.50754E-01$} &
  $0.00102$ \\ \hline
\multirow{5}{*}{\textbf{Class 3}} &
  \multicolumn{1}{c|}{$10$} &
  \multicolumn{1}{c|}{$4.00329E-05$} &
  \multicolumn{1}{c|}{$1.74$} &
  \multicolumn{1}{c|}{$5.49984E-07$} &
  $0.00023$ &
  \multicolumn{1}{c|}{$1$} &
  \multicolumn{1}{c|}{$8.20542E-02$} &
  \multicolumn{1}{c|}{$11.49434$} &
  \multicolumn{1}{c|}{$8.20542E-02$} &
  $0.00034$ \\ \cline{2-11} 
 &
  \multicolumn{1}{c|}{$20$} &
  \multicolumn{1}{c|}{$7.65300E-02$} &
  \multicolumn{1}{c|}{$1.73$} &
  \multicolumn{1}{c|}{$1.62250E-01$} &
  $0.00031$ &
  \multicolumn{1}{c|}{$2$} &
  \multicolumn{1}{c|}{$2.30637E-02$} &
  \multicolumn{1}{c|}{$31.96872$} &
  \multicolumn{1}{c|}{$2.30637E-02$} &
  $0.00051$ \\ \cline{2-11} 
 &
  \multicolumn{1}{c|}{$30$} &
  \multicolumn{1}{c|}{$1.67000E-03$} &
  \multicolumn{1}{c|}{$1.84$} &
  \multicolumn{1}{c|}{$7.95450E-01$} &
  $0.00031$ &
  \multicolumn{1}{c|}{$3$} &
  \multicolumn{1}{c|}{$2.31261E-02$} &
  \multicolumn{1}{c|}{$55.67283$} &
  \multicolumn{1}{c|}{$2.31261E-02$} &
  $0.00072$ \\ \cline{2-11} 
 &
  \multicolumn{1}{c|}{$40$} &
  \multicolumn{1}{c|}{$1.60000E-04$} &
  \multicolumn{1}{c|}{$1.93$} &
  \multicolumn{1}{c|}{$4.63429E+01$} &
  $0.00030$ &
  \multicolumn{1}{c|}{$4$} &
  \multicolumn{1}{c|}{$2.59643E-02$} &
  \multicolumn{1}{c|}{$79.66140$} &
  \multicolumn{1}{c|}{$2.59643E-02$} &
  $0.00081$ \\ \cline{2-11} 
 &
  \multicolumn{1}{c|}{$50$} &
  \multicolumn{1}{c|}{$7.70471E-05$} &
  \multicolumn{1}{c|}{$1.91$} &
  \multicolumn{1}{c|}{$2.28249E+01$} &
  $0.00030$ &
  \multicolumn{1}{c|}{$5$} &
  \multicolumn{1}{c|}{$5.36699E-04$} &
  \multicolumn{1}{c|}{$104.01447$} &
  \multicolumn{1}{c|}{$5.36690E-04$} &
  $0.00094$ \\ \hline
\end{tabular}%
}
\end{center}
\end{table*}

\begin{table}[]
\caption{Source model experiment simulation execution duration for simulation interval: (1) {[}0,0.5{]} seconds (2) {[}0.5,1{]} seconds.}
\label{Table3}
\resizebox{\columnwidth}{!}{%
\begin{tabular}{|c|cc|}
\hline
\multirow{2}{*}{\textbf{Circuit Class}} & \multicolumn{2}{c|}{\textbf{\begin{tabular}[c]{@{}c@{}}Simulation Execution Duration\\ (seconds)\end{tabular}}} \\ \cline{2-3} 
                 & \multicolumn{1}{c|}{\textbf{Simulation Interval 1}} & \textbf{Simulation Interval 2} \\ \hline
\textbf{Class 1} & \multicolumn{1}{c|}{$7.81$}                           & $2.32$                           \\ \hline
\textbf{Class 2} & \multicolumn{1}{c|}{$12.04$}                          & $2.64$                           \\ \hline
\textbf{Class 3} & \multicolumn{1}{c|}{$15.65$}                          & $2.96$                           \\ \hline
\end{tabular}%
}
\end{table}

\begin{table}[]
\caption{Source PIFNN and PINN models hyperparameter optimal settings and count of trainable parameters per circuit class.}
\label{Table4}
\resizebox{\columnwidth}{!}{%
\begin{tabular}{|c|c|c|c|}
\hline
\textbf{Circuit Class} &
  \textbf{Model} &
  \textbf{\begin{tabular}[c]{@{}c@{}}Hyperparameter \\ Optimal Setting\end{tabular}} &
  \textbf{\begin{tabular}[c]{@{}c@{}}Number of \\ Trainable Parameters\end{tabular}} \\ \hline
\multirow{2}{*}{\textbf{Class 1}} & PIFNN & $N_{\text{neurons}} = 10$ & $31$     \\ \cline{2-4} 
                                  & PINN  & $N_{\text{hidden}} = 5$   & $10,351$ \\ \hline
\multirow{2}{*}{\textbf{Class 2}} & PIFNN & $N_{\text{neurons}} = 20$ & $61$     \\ \cline{2-4} 
                                  & PINN  & $N_{\text{hidden}} = 5$  & $10,351$ \\ \hline
\multirow{2}{*}{\textbf{Class 3}} & PIFNN & $N_{\text{neurons}} = 10$ & $31$     \\ \cline{2-4} 
                                  & PINN  & $N_{\text{hidden}} = 5$  & $10,351$ \\ \hline
\end{tabular}%
}
\end{table}

\begin{figure*}[!t]
\centering
\subfloat[]{\includegraphics[width=2.33in]{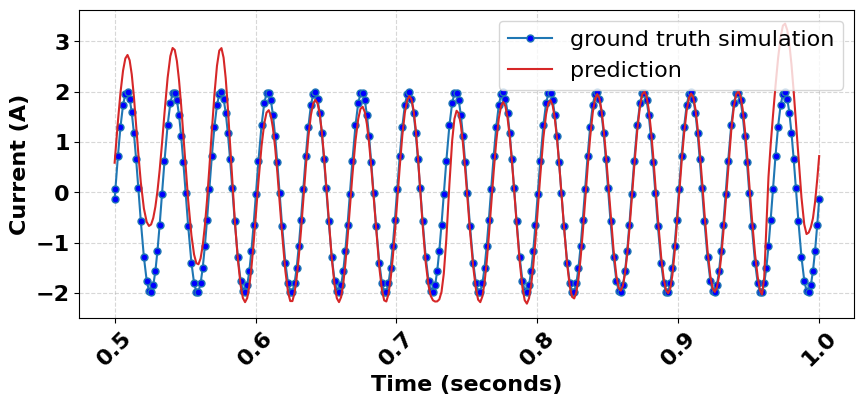}
\label{Fig5_a}}
\hfill
\subfloat[]{\includegraphics[width=2.33in]{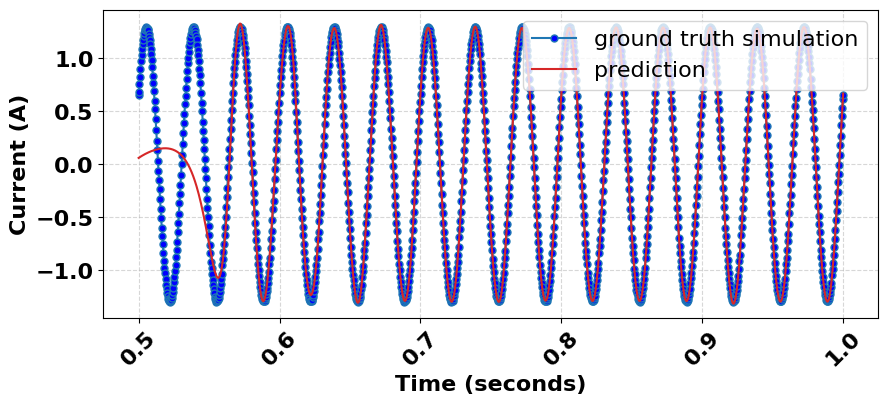}
\label{Fig5_b}}
\hfill
\subfloat[]{\includegraphics[width=2.33in]{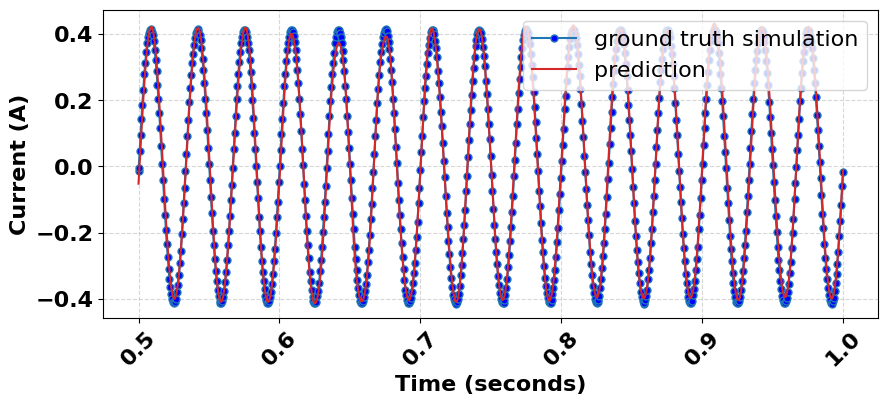}
\label{Fig5_d}}
\vspace{-0.1in}
\subfloat[]{\includegraphics[width=2.33in]{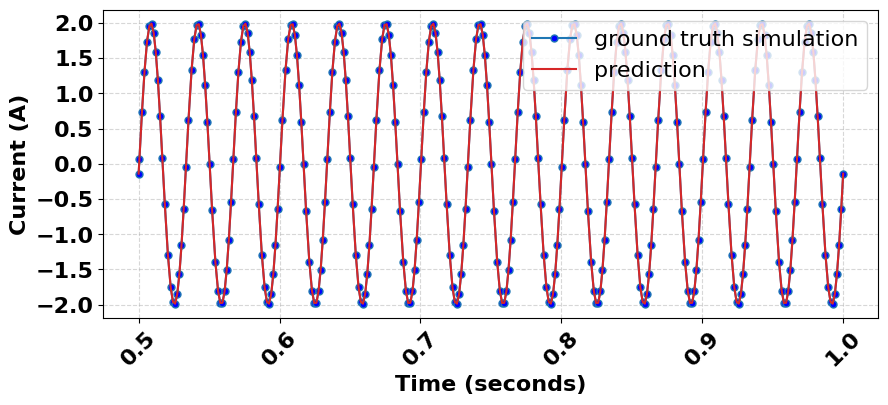}
\label{Fig5_d}}
\hfill
\subfloat[]{\includegraphics[width=2.33in]{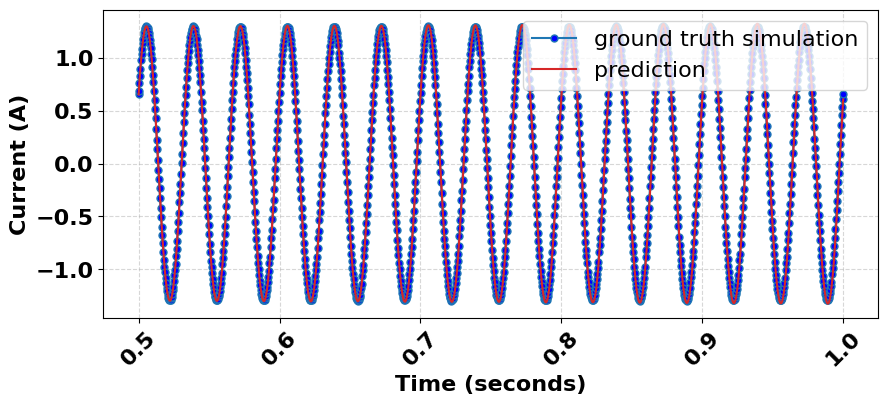}
\label{Fig5_e}}
\hfill
\subfloat[]{\includegraphics[width=2.33in]{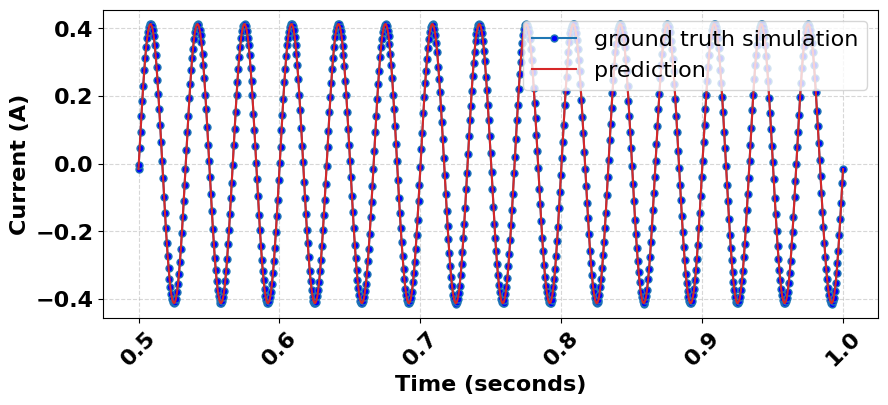}
\label{Fig5_f}}
\caption{Optimal source models performance in testing cases for (a) Class 1 source PINN model, (b) Class 2 source PINN model, (c) Class 3 source PINN model, (d) Class 1 source PIFNN model, (e) Class 2 source PIFNN model, and (f) Class 3 source PIFNN model.}
\label{Fig5}
\end{figure*}

For each circuit class, the optimal source PIFNN and PINN models are selected based on the lowest testing MSE from performance results reported in Table~\ref{Table2}. The corresponding configurations and number of trainable parameters per hyperparameter setting are summarized in Table~\ref{Table4}. The performance of these optimal models, compared to the actual simulation, is illustrated in Fig.~\ref{Fig5}.


Table~\ref{Table5} presents the Wilcoxon sum-rank test results comparing prediction errors of the optimal PIFNN and PINN models. A $P$-value threshold of $0.05$ is used, with lower values indicating statistical significance. The results confirm that PIFNN predictions are statistically significant across circuit classes 1, 2, and 3.

\begin{table}[]
\caption{Wilcoxon sum rank significance test for PIFNN and PINN testing prediction errors.}
\label{Table5}
\resizebox{\columnwidth}{!}{%
\begin{tabular}{|c|c|c|c|}
\hline
\textbf{Circuit Class} &
  \textbf{\begin{tabular}[c]{@{}c@{}}Wilcoxon \\ Rank-sum \\ Statistic\end{tabular}} &
  \textbf{$P$-value} &
  \multicolumn{1}{l|}{\textbf{Prediction Errors Difference}} \\ \hline
\textbf{Class 1} & $3.36607$  & $7.62500E-04$ & Statistically Significant \\ \hline
\textbf{Class 2} & $4.61310$  & $3.96706E-06$ & Statistically Significant \\ \hline
\textbf{Class 3} & $-11.74682$ & $7.33334E-32$ & Statistically Significant \\ \hline
\end{tabular}%
}
\end{table}

\subsection{Electrical Circuits Analysis Experiment}

\begin{table}[]
\caption{New circuit parameter values per circuit class.}
\label{Table6}
\resizebox{\columnwidth}{!}{%
\begin{tabular}{|>{\centering\arraybackslash}p{0.12\linewidth}|>{\centering\arraybackslash}p{0.11\linewidth}|>{\centering\arraybackslash}p{0.11\linewidth}|>{\centering\arraybackslash}p{0.11\linewidth}|>{\centering\arraybackslash}p{0.12\linewidth}|>{\centering\arraybackslash}p{0.12\linewidth}|}
\hline
\multicolumn{1}{|l|}{\textbf{Circuit Class}} & $R$ (Ohms) & $L$ (Henry) & $C$ (Farads) & $V_{\max}$ (Volts) & $f$ (Hertz) \\ \hline
\textbf{Class 1} & $10$ & $0.01$  & $0.0009$ & $15$  & $25$ \\ \hline
\textbf{Class 2} & $10$ & $0.0009$ & $0.00006$  & $90$ & $40$ \\ \hline
\textbf{Class 3} & $25$ & $0.009$  & $0.000065$ & $100$ & $60$ \\ \hline
\end{tabular}%
}
\end{table}

In the second experiment, the objective is to efficiently and effectively approximate the load current for new circuit parameter values summarized in Table~\ref{Table6}. Circuit analysis is performed by randomly changing all circuit parameters per class, and the updated physics loss terms guide the learning process to converge to the new load current solutions in an unsupervised (label-free) manner.

This experiment evaluates the impact of intra-circuit class TL by alternating between activating and deactivating TL during unsupervised training. When TL is active, target PIFNN and PINN models are fine-tuned; when inactive, they are trained from scratch. Target models maintain the same architecture as their source counterparts, including input, hidden, and output neurons and hidden layers. Training/Fine-tuning is guided solely by the physics loss term updated with the new circuit parameters, while ground-truth simulation data is used only to assess performance in terms of MSE and computational efficiency.

The number of training epochs for both PIFNN and PINN models is reduced to half to improve computational efficiency while maintaining sufficient training time to achieve convergence. For the PIFNN models, a relatively small number of epochs ($300$) is chosen due to their ability to leverage pre-trained parameters effectively, requiring less training time for fine-tuning. The learning rate for PIFNN models is set to $0.1$ to enable moderately large updates during fine-tuning, striking a balance between rapid convergence and stability. In contrast, PINN models are trained for a significantly higher number of epochs ($50,000$) with the same learning rate ($0.001$). The fine-tuning/training stages are performed for the temporal interval $t = [0, 0.5]$, and the testing is performed on the temporal interval $t = [0.5, 1]$.

The results of this experiment for all circuit classes are summarized in Table~\ref{Table7}, and the performance of testing cases for circuit class 3 for PIFNN and PINN models, in comparison to the ground-truth simulations, is illustrated in Fig.~\ref{Fig6}. At this point, the reported simulation re-run execution duration times reported in Table~\ref{Table8} represent a fair point of efficiency comparison with PINN and PIFNN training, fine-tuning, or testing duration times, as they reflect the simulation time required to re-solve the system of DEs for new circuit parameters without accounting for equations derivation time.
\begin{table*}
\begin{center}
\caption{Simulation results comparison with PIFNN and PINN unsupervised training and fine-tuning for circuit analysis tasks.}
\label{Table7}
\resizebox{\textwidth}{!}{%
\begin{tabular}{|c|c|ccccc|ccccc|}
\hline
\multirow{3}{*}{\textbf{Circuit Class}} &
  \multirow{3}{*}{\textbf{Transfer Learning}} &
  \multicolumn{5}{c|}{\textbf{PIFNN}} &
  \multicolumn{5}{c|}{\textbf{PINN}} \\ \cline{3-12} 
 &
   &
  \multicolumn{1}{c|}{\multirow{2}{*}{\textbf{$N_{\text{neurons}}$}}} &
  \multicolumn{2}{c|}{\textbf{Training}} &
  \multicolumn{2}{c|}{\textbf{Testing}} &
  \multicolumn{1}{c|}{\multirow{2}{*}{\textbf{$N_{\text{hidden}}$}}} &
  \multicolumn{2}{c|}{\textbf{Training}} &
  \multicolumn{2}{c|}{\textbf{Testing}} \\ \cline{4-7} \cline{9-12} 
 &
   &
  \multicolumn{1}{c|}{} &
  \multicolumn{1}{c|}{\textbf{MSE}} &
  \multicolumn{1}{c|}{\textbf{\begin{tabular}[c]{@{}c@{}}Duration \\ (seconds)\end{tabular}}} &
  \multicolumn{1}{c|}{\textbf{MSE}} &
  \textbf{\begin{tabular}[c]{@{}c@{}}Duration\\ (seconds)\end{tabular}} &
  \multicolumn{1}{c|}{} &
  \multicolumn{1}{c|}{\textbf{MSE}} &
  \multicolumn{1}{c|}{\textbf{\begin{tabular}[c]{@{}c@{}}Duration \\ (minutes)\end{tabular}}} &
  \multicolumn{1}{c|}{\textbf{MSE}} &
  \textbf{\begin{tabular}[c]{@{}c@{}}Duration \\ (seconds)\end{tabular}} \\ \hline
\multirow{2}{*}{\textbf{Class 1}} &
  \textbf{Deactivated} &
  \multicolumn{1}{c|}{\multirow{2}{*}{$10$}} &
  \multicolumn{1}{c|}{$8.49883E-01$} &
  \multicolumn{1}{c|}{$1.18$} &
  \multicolumn{1}{c|}{$2.97940E+00$} &
  $0.00024$ &
  \multicolumn{1}{c|}{\multirow{2}{*}{$5$}} &
  \multicolumn{1}{c|}{$4.31025E-01$} &
  \multicolumn{1}{c|}{$8.99758$} &
  \multicolumn{1}{c|}{\textbf{$3.24873E-01$}} &
  0.00154 \\ \cline{2-2} \cline{4-7} \cline{9-12} 
 &
  \textbf{Activated} &
  \multicolumn{1}{c|}{} &
  \multicolumn{1}{c|}{\textbf{$1.39395E-02$}} &
  \multicolumn{1}{c|}{$1.16$} &
  \multicolumn{1}{c|}{\textbf{$1.33725E-03$}} &
  $0.00028$ &
  \multicolumn{1}{c|}{} &
  \multicolumn{1}{c|}{\textbf{$3.39741E-01$}} &
  \multicolumn{1}{c|}{$8.93133$} &
  \multicolumn{1}{c|}{$4.04398E-01$} &
  $0.00074$ \\ \hline
\multirow{2}{*}{\textbf{Class 2}} &
  \textbf{Deactivated} &
  \multicolumn{1}{c|}{\multirow{2}{*}{$20$}} &
  \multicolumn{1}{c|}{$2.83410E-01$} &
  \multicolumn{1}{c|}{$1.03$} &
  \multicolumn{1}{c|}{$8.03380E+00$} &
  $0.00040$ &
  \multicolumn{1}{c|}{\multirow{2}{*}{$5$}} &
  \multicolumn{1}{c|}{$1.70639E-01$} &
  \multicolumn{1}{c|}{$28.23500$} &
  \multicolumn{1}{c|}{$8.64462E-01$} &
  $0.00274$ \\ \cline{2-2} \cline{4-7} \cline{9-12} 
 &
  \textbf{Activated} &
  \multicolumn{1}{c|}{} &
  \multicolumn{1}{c|}{\textbf{$7.63000E-03$}} &
  \multicolumn{1}{c|}{$1.26$} &
  \multicolumn{1}{c|}{\textbf{$2.20630E-06$}} &
  $0.00048$ &
  \multicolumn{1}{c|}{} &
  \multicolumn{1}{c|}{\textbf{$9.11557E-02$}} &
  \multicolumn{1}{c|}{$32.69000$} &
  \multicolumn{1}{c|}{\textbf{$9.11567E-02$}} &
  $0.00143$ \\ \hline
\multirow{2}{*}{\textbf{Class 3}} &
  \textbf{Deactivated} &
  \multicolumn{1}{c|}{\multirow{2}{*}{$10$}} &
  \multicolumn{1}{c|}{$1.24646E+01$} &
  \multicolumn{1}{c|}{$1.01$} &
  \multicolumn{1}{c|}{$1.34512E+01$} &
  $0.00031$ &
  \multicolumn{1}{c|}{\multirow{2}{*}{$5$}} &
  \multicolumn{1}{c|}{$8.49827E+00$} &
  \multicolumn{1}{c|}{$43.57834$} &
  \multicolumn{1}{c|}{$8.49827E+00$} &
  $0.00087$ \\ \cline{2-2} \cline{4-7} \cline{9-12} 
 &
  \textbf{Activated} &
  \multicolumn{1}{c|}{} &
  \multicolumn{1}{c|}{\textbf{$4.17143E-02$}} &
  \multicolumn{1}{c|}{$1.15$} &
  \multicolumn{1}{c|}{\textbf{$5.95619E-03$}} &
  $0.00267$ &
  \multicolumn{1}{c|}{} &
  \multicolumn{1}{c|}{\textbf{$1.86097E+00$}} &
  \multicolumn{1}{c|}{$45.65167$} &
  \multicolumn{1}{c|}{\textbf{$1.86097E+00$}} &
  $0.00136$ \\ \hline
\end{tabular}
}
\end{center}
\end{table*}

\begin{table}[]
\caption{Target model experiment simulation execution duration for simulation interval (1) {[}0,0.5{]} seconds (2) {[}0.5,1{]} seconds.}
\label{Table8}
\resizebox{\columnwidth}{!}{%
\begin{tabular}{|c|cc|}
\hline
\multirow{2}{*}{\textbf{Circuit Class}} & \multicolumn{2}{c|}{\textbf{\begin{tabular}[c]{@{}c@{}}Simulation Execution Duration\\ (seconds)\end{tabular}}} \\ \cline{2-3} 
                 & \multicolumn{1}{c|}{\textbf{Simulation Interval 1}} & \textbf{Simulation Interval 2} \\ \hline
\textbf{Class 1}  & \multicolumn{1}{c|}{$3.22$}                           & $2.33$                           \\ \hline
\textbf{Class 2} & \multicolumn{1}{c|}{$3.76$}                           & $2.65$                         \\ \hline
\textbf{Class 3} & \multicolumn{1}{c|}{$3.84$}                           & $2.89$                           \\ \hline
\end{tabular}%
}
\end{table}

\renewcommand{\figVscale}{0.7}
\begin{figure*}[!t]
\centering
\subfloat[]{\scalebox{1}[\figVscale]{\includegraphics[width=3.5in]{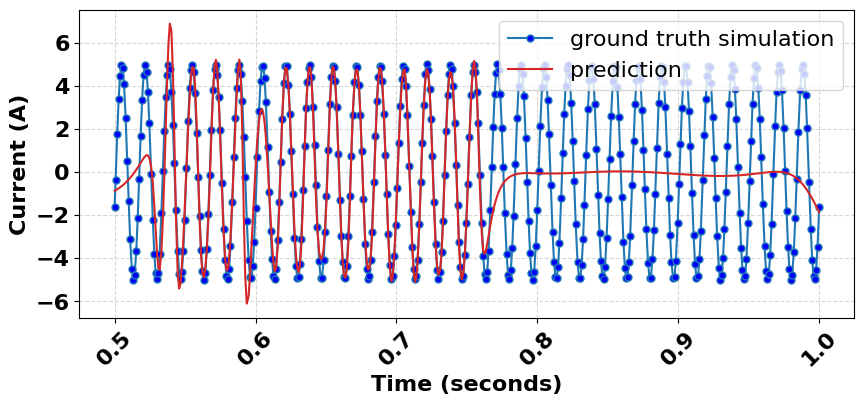}}\label{Fig6_a}}%
\hfill
\subfloat[]{\scalebox{1}[\figVscale]{\includegraphics[width=3.5in]{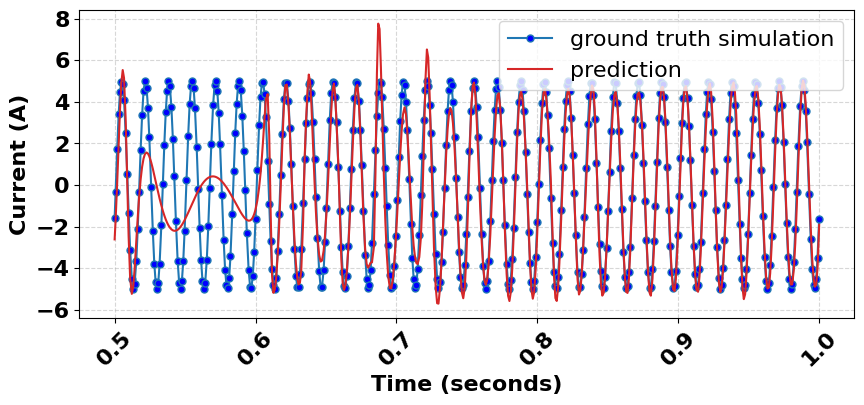}}\label{Fig6_b}}%
\vspace{-0.5\baselineskip} 
\subfloat[]{\scalebox{1}[\figVscale]{\includegraphics[width=3.5in]{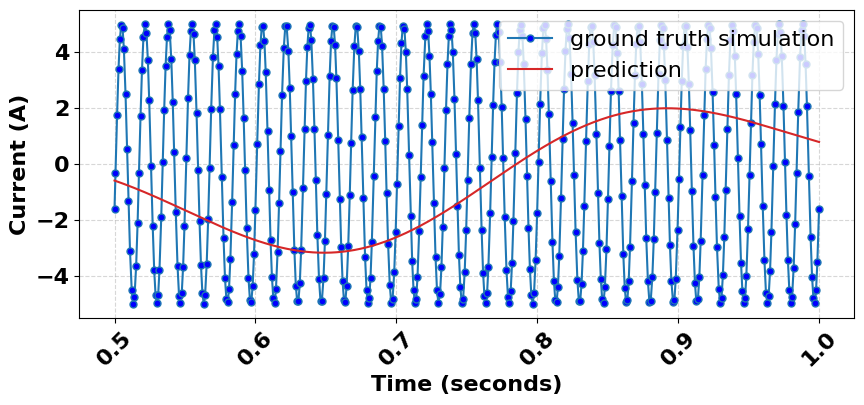}}\label{Fig6_c}}%
\hfill
\subfloat[]{\scalebox{1}[\figVscale]{\includegraphics[width=3.5in]{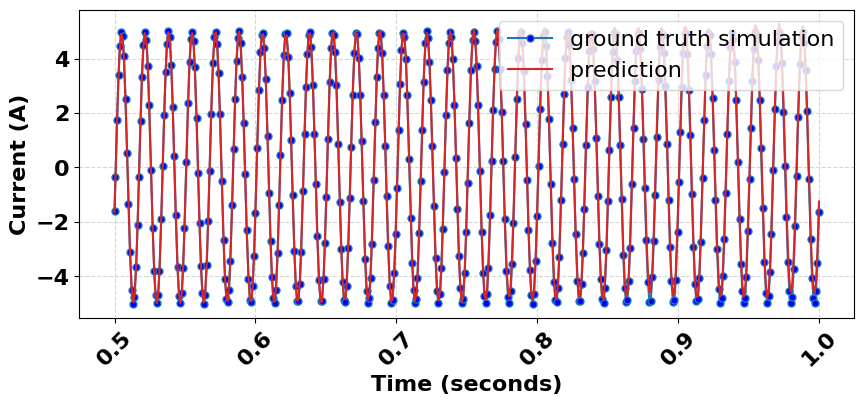}}\label{Fig6_d}}%
\caption{Impact of TL for circuit class 3 analysis: (a) PINN TL deactivated, (b) PINN TL activated, (c) PIFNN TL deactivated, and (d) PIFNN TL activated.}
\label{Fig6}
\end{figure*}

Furthermore, we conduct the Wilcoxon sum rank significance for the circuit analysis experiment prediction errors and report the test results in Table~\ref{Table9}. The significance test is conducted to verify that the proposed PIFNN model predictions are statistically significant when compared to the PINN model predictions with TL activated. 

\begin{table}[]
\caption{Wilcoxon sum rank significance test for PIFNN and PINN testing prediction errors when TL is activated.}
\label{Table9}
\resizebox{\columnwidth}{!}{%
\begin{tabular}{|c|c|c|c|}
\hline
\textbf{Circuit Class} & \textbf{Test Statistic} & \textbf{$P$-value} & \multicolumn{1}{l|}{\textbf{Prediction Errors Difference}} \\ \hline
\textbf{Class 1} & $7.82871$  & $4.92920E-15$ & Statistically Significant \\ \hline
\textbf{Class 2} & $2.25793$  & $2.39502E-02$ & Statistically Significant \\ \hline
\textbf{Class 3} & $10.35748$ & $3.87023E-25$ & Statistically Significant \\ \hline
\end{tabular}%
}
\end{table}

\subsection{Generalization Experiment}

After conducting the previous electrical circuits analysis experiment and studying the impact of intra-circuit class TL, we design a generalization experiment to verify the generalizability of the G-PIFNN framework by performing a cross-circuit class TL process. The goal of this experiment is to examine the generalizability of our framework in analyzing new, unprecedented circuit topologies in an unsupervised manner. Given the observations from the previous experiment, we are motivated to proceed with a generalization experiment for the PIFNN models only.

In this experiment, the cross-circuit class TL process spans multiple circuit classes, involving parameter exchanges between distinct circuit classes. Source models pre-trained for one class are utilized for electric circuits analysis tasks for the other two circuit classes. Therefore, from each pre-trained source model, we expand two target models. For example, a circuit class 1 pre-trained source PIFNN model is harnessed to fine-tune two target PIFNN models, which are linked with circuit classes  2 and 3 target learning tasks. The learning tasks linked with these target models are similar to the ones described in the previous experiment. Specifically, approximating an accurate solution of the load current in each associated target circuit, given the set of circuit parameters per target circuit class summarized in Table~\ref{Table6}.

\begin{figure*}[!t]
\centering
\subfloat[]{\includegraphics[width=2.3in]{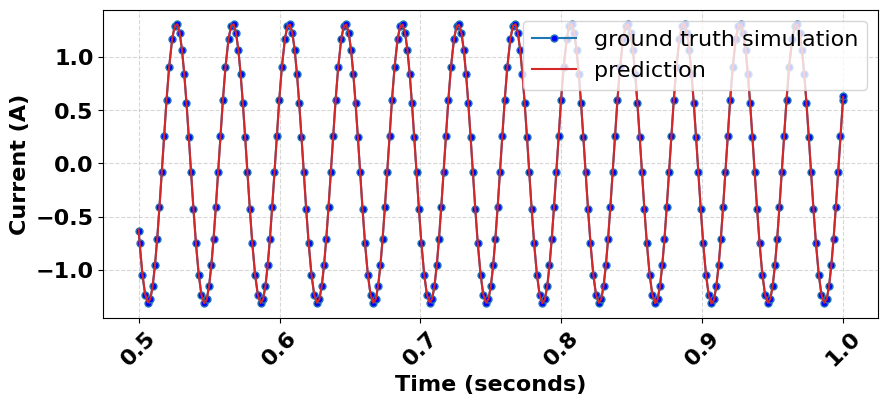}
\label{Fig7_a}}
\hfill
\subfloat[]{\includegraphics[width=2.3in]{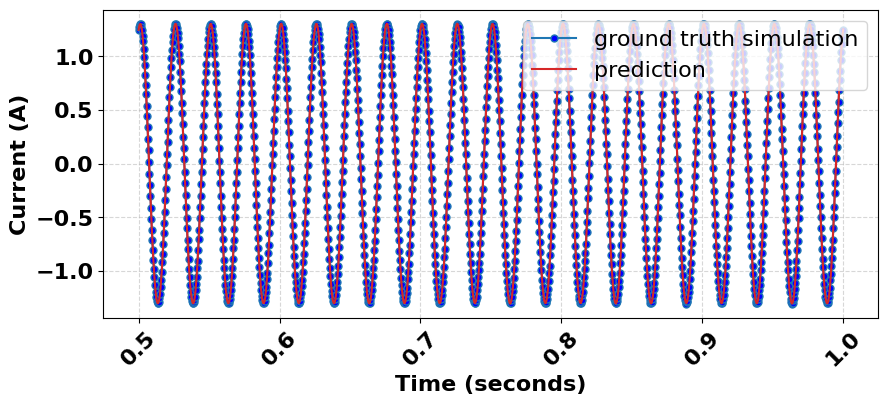}
\label{Fig7_b}}
\hfill
\subfloat[]{\includegraphics[width=2.3in]{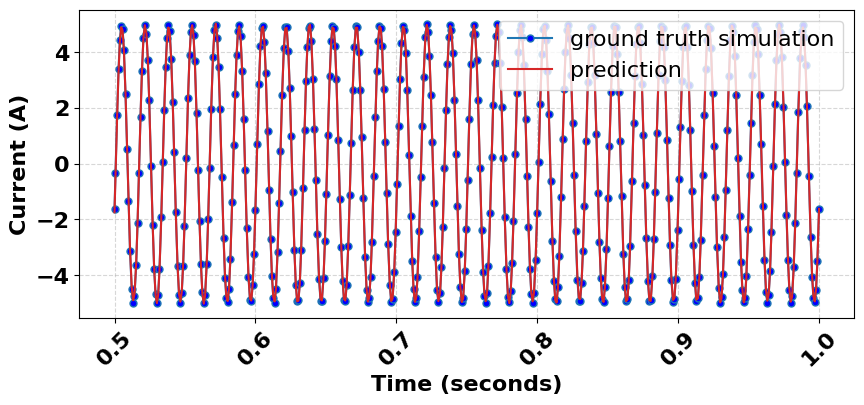}
\label{Fig7_c}}
\vspace{-0.1in}
\subfloat[]{\includegraphics[width=2.3in]{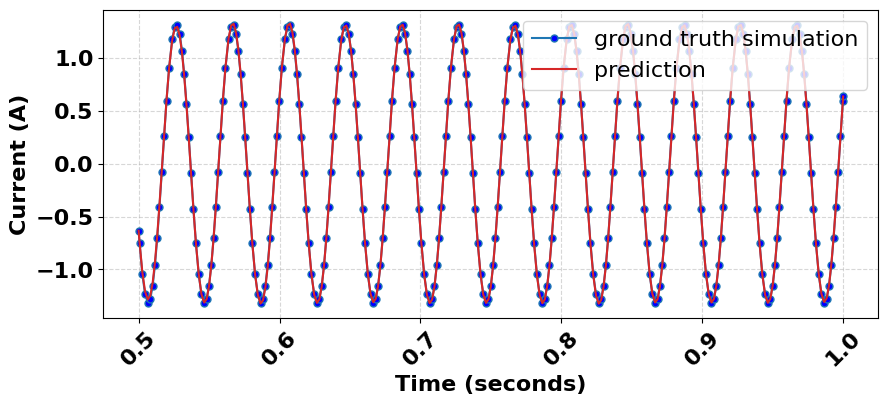}
\label{Fig7_d}}
\hfill
\subfloat[]{\includegraphics[width=2.3in]{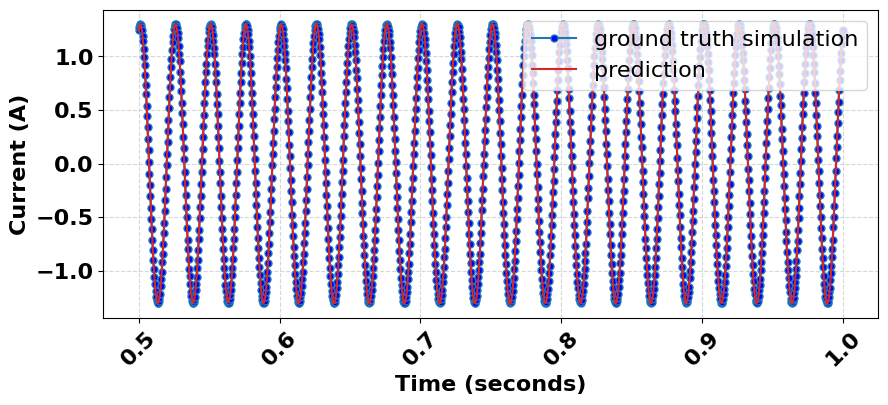}
\label{Fig7_e}}
\hfill
\subfloat[]{\includegraphics[width=2.3in]{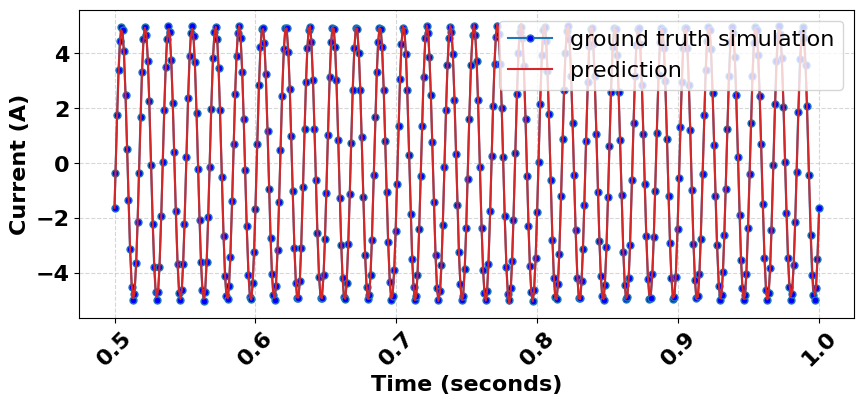}
\label{Fig7_f}}
\caption{Testing case illustrations of generalization experiment for (a) $T_1\_S_2$, (b) $T_2\_S_1$, (c) $T_3\_S_1$, (d) $T_1\_S_3$, (e) $T_2\_S_3$, and (f) $T_3\_S_2$.}
\label{Fig7}
\end{figure*}

The number of epochs and learning rate settings for this experiment are the same as in the previous experiment. Detailed performance results of the generalization experiment are reported in Table~\ref{Table10}. Furthermore, illustrations of the testing cases are showcased in Fig.~\ref{Fig7}. The notation $T_1\_S_2$ in Fig.~\ref{Fig7_a} is introduced to indicate that the target model of circuit class 1 was fine-tuned on the source model parameters of circuit class 2.
\begin{table*}[]
\caption{PIFNN generalization experiment results.}
\label{Table10}
\resizebox{\textwidth}{!}{%
\begin{tabular}{|c|ccc|ccc|ccc|}
\hline
\multirow{2}{*}{\textbf{Circuit Class}} &
  \multicolumn{3}{c|}{\textbf{\begin{tabular}[c]{@{}c@{}}Target \\ Class 1\end{tabular}}} &
  \multicolumn{3}{c|}{\textbf{\begin{tabular}[c]{@{}c@{}}Target \\ Class 2\end{tabular}}} &
  \multicolumn{3}{c|}{\textbf{\begin{tabular}[c]{@{}c@{}}Target \\ Class 3\end{tabular}}} \\ \cline{2-10} 
 &
  \multicolumn{1}{c|}{\textbf{\begin{tabular}[c]{@{}c@{}}Fine-Tuning \\ MSE\end{tabular}}} &
  \multicolumn{1}{c|}{\textbf{\begin{tabular}[c]{@{}c@{}}Testing \\ MSE\end{tabular}}} &
  \textbf{\begin{tabular}[c]{@{}c@{}}Fine-Tuning \\ Duration \\ (seconds)\end{tabular}} &
  \multicolumn{1}{c|}{\textbf{\begin{tabular}[c]{@{}c@{}}Fine-Tuning \\ MSE\end{tabular}}} &
  \multicolumn{1}{c|}{\textbf{\begin{tabular}[c]{@{}c@{}}Testing \\ MSE\end{tabular}}} &
  \textbf{\begin{tabular}[c]{@{}c@{}}Fine-Tuning \\ Duration \\ (seconds)\end{tabular}} &
  \multicolumn{1}{c|}{\textbf{\begin{tabular}[c]{@{}c@{}}Fine-Tuning\\ MSE\end{tabular}}} &
  \multicolumn{1}{c|}{\textbf{\begin{tabular}[c]{@{}c@{}}Testing \\ MSE\end{tabular}}} &
  \textbf{\begin{tabular}[c]{@{}c@{}}Fine-Tuning \\ Duration \\ (seconds)\end{tabular}} \\ \hline
\textbf{\begin{tabular}[c]{@{}c@{}}Source \\ Class 1\end{tabular}} &
  \multicolumn{3}{c|}{\text{reported in Table~\ref{Table7}}} &
  \multicolumn{1}{c|}{$7.63752E-03$} &
  \multicolumn{1}{c|}{\textbf{$1.07890E-06$}} &
  $1.60$ &
  \multicolumn{1}{c|}{$1.48302E-01$} &
  \multicolumn{1}{c|}{\textbf{$3.56978E-03$}} &
  $1.58$ \\ \hline
\textbf{\begin{tabular}[c]{@{}c@{}}Source \\ Class 2\end{tabular}} &
  \multicolumn{1}{c|}{$1.38968E-02$} &
  \multicolumn{1}{c|}{$5.59506E-03$} &
 $1.16$ &
  \multicolumn{3}{c|}{\text{reported in Table~\ref{Table7}}} &
  \multicolumn{1}{c|}{$1.03533E+00$} &
  \multicolumn{1}{c|}{$2.43015E-02$} &
  $1.35$ \\ \hline
\textbf{\begin{tabular}[c]{@{}c@{}}Source \\ Class 3\end{tabular}} &
  \multicolumn{1}{c|}{$4.40398E-02$} &
  \multicolumn{1}{c|}{\textbf{$6.53709E-05$}} &
  $1.28$ &
  \multicolumn{1}{c|}{$1.56925E-02$} &
  \multicolumn{1}{c|}{$9.92714E-06$} &
  $1.63$ &
  \multicolumn{3}{c|}{\text{reported in Table~\ref{Table7}}} \\ \hline
\end{tabular}%
}
\end{table*}

\subsection{Results Discussion}

In this section, we present a collective analysis and summary of the results obtained from all conducted experiments in this paper. By examining the outcomes of each experiment, we aim to highlight the strengths and limitations of the proposed framework.

In the first experiment, the PIFNN model consistently outperformed the standard PINN across all training and testing scenarios in approximating solutions for low-order to high-order DEs associated with electric circuits. As reported in Tables~\ref{Table2}-\ref{Table3}, simulation time increases with DE order, but PIFNN achieved notably faster testing inference time (on average $0.00033$ s) compared to re-running simulations (on average $2.64$ s) for testing intervals. Furthermore, PIFNN demonstrated superior effectiveness in both low MSE magnitudes and signal reconstruction as illustrated in Fig.~\ref{Fig5}, effectively capturing system dynamics and signal periodicity after only $600$ epochs. In contrast, PINN failed to generalize to unseen data even after $100,000$ epochs, regardless of network depth ($N_{\text{layers}}$).

In the second experiment, integrating intra-circuit class TL within the G-PIFNN framework significantly enhances the performance of electrical circuit analysis. The unsupervised fine-tuning approach enabled PIFNN models to effectively replicate ground-truth signals for all circuit classes, as shown specifically for circuit class 3 in Fig.\ref{Fig6}. PIFNN consistently achieved lower MSEs in both fine-tuning and testing, with substantial efficiency gains over simulation software: $67.31\%$ in fine-tuning and $99.96\%$ in testing, on average for all three circuit classes. Compared to target PINN models, fine-tuned PIFNNs required fewer trainable parameters, epochs, and fine-tuning time, while delivering superior predictive performance across all circuit classes.

In the final experiment, G-PIFNN generalized across circuit topologies without requiring new simulations or ground-truth data. As Table~\ref{Table10} and Fig.~\ref{Fig7} demonstrate, cross-circuit TL effectively replicates signals, yielding average MSEs of $5.59\times10^{-3}$ (fine-tuning) and $5.99\times10^{-3}$ (testing) across all classes. Notably, in specific cases per circuit class, cross-circuit class TL significantly outperforms intra-circuit class TL—for target class~1 using class~3 as source, testing $\text{MSE}=6.53709\times10^{-5}$ vs.\ $\text{MSE}=1.33725\times10^{-3}$ achieved by the counterpart target model via intra-circuit class TL. Cross-circuit TL also enables a fair compute comparison with simulation durations in Table~\ref{Table8}: for class~1, fine-tuning via cross-circuit TL averaged $1.22$ seconds compared to simulation duration\ $3.22$ seconds (62.11\% faster), with similar gains for circuit classes~2 and~3 respectively (57.18\%, 61.72\%).

\section{Conclusion and Future Research}\label{sec: conclusion}

This paper introduces G-PIFNN, a novel framework for circuit analysis based on a physically enhanced PINN architecture. The enhancement integrates a PAF and restructures the network into a PIFNN, yielding a lightweight, interpretable, and scalable modeling approach for electrical systems. G-PIFNN operates in three stages: pre-processing, source model training, and target model fine-tuning. In pre-processing, governing DEs are derived via the BG method as a basis for generalization. In source training, a PIFNN is pre-trained for each topology to capture distinct dynamics, while target fine-tuning adapts these models to new parameter settings. Two TL strategies are employed: intra-circuit class TL for parameter variations and cross-circuit class TL for topology generalization, both performed in an unsupervised (label-free) manner. Experiments confirm G-PIFNN’s effectiveness, demonstrating lower complexity, improved efficiency, and strong generalizability across varying circuit topologies.

Furthermore, the PINN's physical enhancement thought process introduced in this paper can be applied to other scientific fields outside the electrical domain. Future research directions include learning specialized PAFs that can handle transient system states and power electronics switching behaviors through the use of Kormogolov-Arnold Networks (KANs) \cite{KANs}. Additionally, another possible direction is utilizing the effectiveness and efficiency introduced by G-PIFNN in developing an advanced inverse learning algorithm for circuit design tasks. At last, the efficacy and generalization brought by G-PIFNN can be seamlessly integrated in realizing high-fidelity and interpretable Digital Twins of complex power grids and power electronics devices.

\bibliographystyle{IEEEtran}
\bibliography{bibfile}

\begin{IEEEbiography}[{\includegraphics[width=1in,height=1.25in,clip,keepaspectratio]{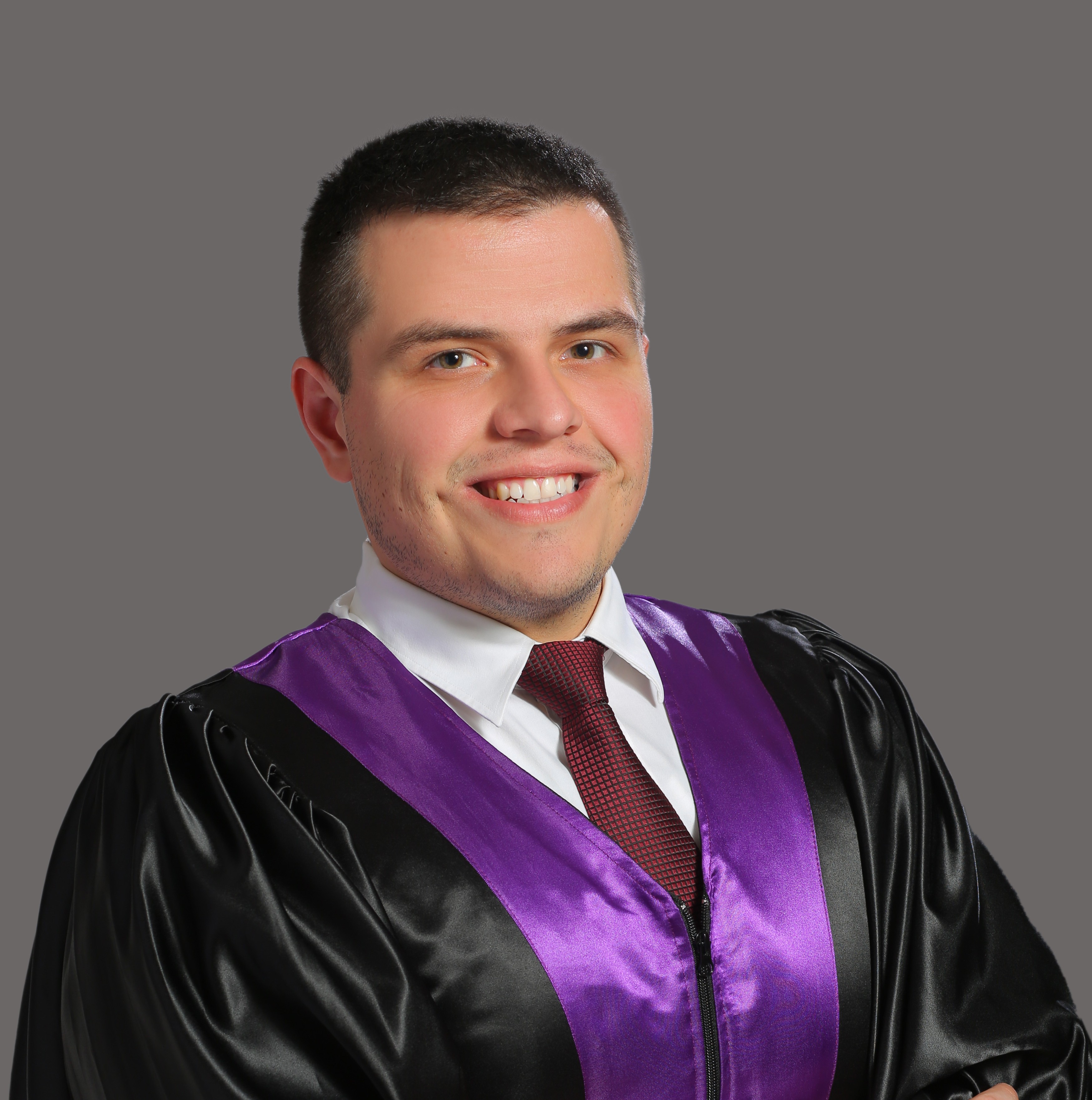}}]
{Ibrahim Shahbaz} (Student Member, IEEE) received the B.Sc. degree in electrical engineering from the University of Jordan, Amman, Jordan, and the M.Sc. degree in Data Science from Princess Sumaya University for Technology, Amman, Jordan. He is currently a Ph.D. student in electrical engineering at Texas A\&M University, College Station, TX, USA. His research interests at the iSTAR Lab include intelligent cyber-physical systems, resilient smart grids, and scientific machine learning.
\end{IEEEbiography}

\begin{IEEEbiography}[{\includegraphics[width=1in,height=1.25in,clip,keepaspectratio]{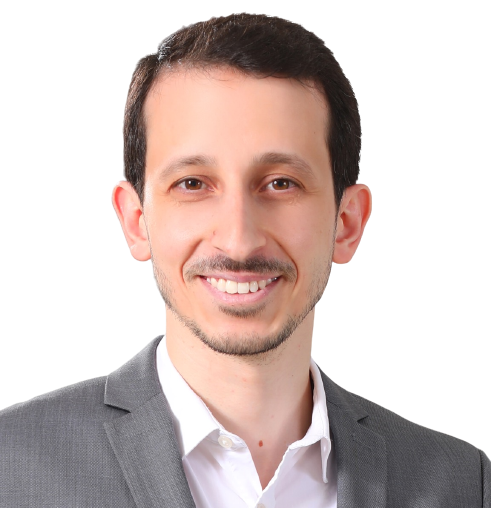}}]{MOHAMMAD J. ABDEL-RAHMAN} (Senior~Member, IEEE) received the Ph.D. degree from the Electrical and Computer Engineering (ECE) Department at the University of Arizona, Tucson, AZ, USA. He is currently an Associate Professor with the Data Science Department at Princess Sumaya University for Technology, Amman, Jordan, and an Adjunct Assistant Professor with the ECE Department at Virginia Tech, Blacksburg, VA, USA. His current research lies at the intersection of artificial intelligence and operations research, with applications in communications, networking, energy, healthcare, and transportation.

Dr. Abdel-Rahman is a member of the Global Young Academy. He currently serves as an Associate Editor for \textsc{IEEE Transactions on Cognitive Communications and Networking}, \textsc{IEEE Access}, and \textit{Wireless Personal Communications} (Springer Nature). He regularly contributes to the technical program committees of major international conferences.
\end{IEEEbiography}

\begin{IEEEbiography}[{\includegraphics[width=1in,height=1.25in,clip,keepaspectratio]{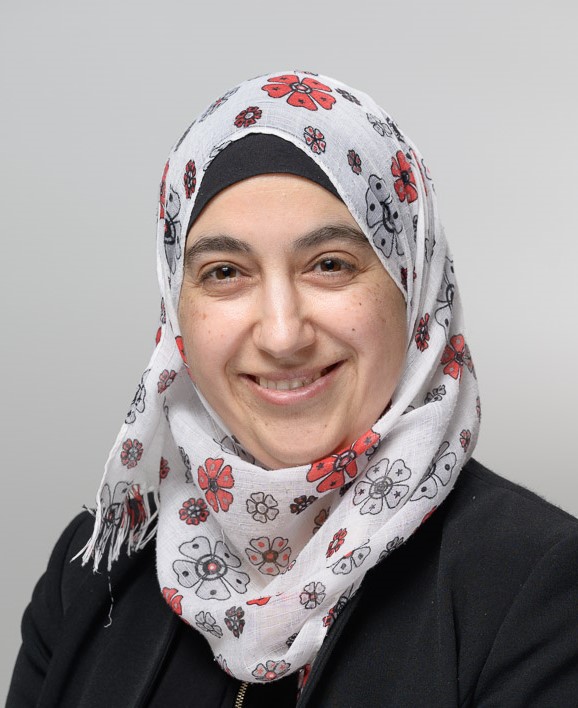}}]{Eman Hammad} (Senior~Member, IEEE) PhD, is an assistant professor leading the innovations in systems trust and resilience lab (iSTAR) at Texas A\&M University. She is also the director of the Texas A\&M Data Institute Thematic Lab - SPARTA Security, Privacy and Trust for AI. Hammad received her PhD from the Electrical and Computer Engineering Department at the University of Toronto (UofT). Hammad's research interests include large-scale adaptive, reliable, and trustworthy heterogeneous networks, connected intelligence, systems' integration and interoperability, metrics-informed design, security and resilience by design.
\end{IEEEbiography}

\end{document}